\documentclass[12pt,superscriptaddress,showpacs]{revtex4}
\usepackage{epsfig}
\usepackage{bm}
\usepackage{amsmath}

\begin{document}

\title{Transient backbending behaviour in the Ising model with fixed 
magnetization}
\author{F. Gulminelli}
\affiliation{LPC Caen (IN2P3-CNRS/ISMRA et Universit\'{e}), F-14050
Caen C\'{e}dex, France}
\author{J.M. Carmona}
\affiliation{Departamento de F\'{i}sica Te\'{o}rica, Universidad de Zaragoza, 
50009 Zaragoza, Spain}
\author{Ph. Chomaz}
\affiliation{GANIL (DSM-CEA/IN2P3-CNRS), B.P.5027, F-14076 Caen C\'{e}dex 5, 
France}
\author{J. Richert}
\affiliation{Laboratoire de Physique Th\'{e}orique (UMR7085), 
Universit\'e Louis Pasteur, 3 rue de l'Universit\'{e}, 
67084 Strasbourg C\'{e}dex, France}
\author{S. Jim\'{e}nez}
\affiliation{Departamento de F\'{i}sica Te\'{o}rica, Universidad de Zaragoza, 
50009 Zaragoza, Spain}
\author{V.Regnard}
\affiliation{LPC Caen (IN2P3-CNRS/ISMRA et Universit\'{e}), F-14050
Caen C\'{e}dex, France}
\affiliation{GANIL (DSM-CEA/IN2P3-CNRS), B.P.5027, F-14076 Caen C\'{e}dex 5, 
France}

\begin{abstract}
{The physical origin of the backbendings in the equations of state of
finite but not necessarily small systems is studied in the Ising model with
fixed magnetization (IMFM) by means of the topological properties of the
observable distributions and the analysis of the largest cluster with
increasing lattice size. Looking at the convexity anomalies of the IMFM
thermodynamic potential, it is shown that the order of the transition at the
thermodynamic limit can be recognized in finite systems independently of the
lattice size. General statistical mechanics arguments and analytical
calculations suggest that the backbending in the caloric curve is a
transient behaviour which should not converge to a plateau in the
thermodynamic limit, while the first order transition is signalled by a
discontinuity in other observables.}
\end{abstract}

\maketitle

\section{Introduction}
\label{sec1}

The %presence 
origin of singularities in thermodynamic functions which characterize
infinite systems undergoing phase transitions is a central issue in
statistical mechanics. %A central issue in statistical mechanics is to
%understand how the analytic partition sum of a finite size system 
%acquires a singularity in the thermodynamic limit if the system 
%undergoes a phase transition. 
A possible approach deals with the distribution of zeroes of the partition
sum in the complex temperature plane~\cite{grossman}. Alternatively it has
been observed that non-analyticities can originate from a backbending in an
equation of state corresponding to an anomalous curvature of the
thermodynamic potential surface~\cite{prl99}. In particular a convex
intruder in the microcanonical entropy, leading to a backbending in the
functional relationship between temperature and energy (caloric curve) has
been proposed as a general definition of first order phase transitions in
finite systems~\cite{gross}. Indeed negative heat capacities have been
observed and connected to  first order phase transitions in
different models~\cite{iso-iso,ruffo,cohen} and even experimentally 
measured~\cite{michela,haberland,farizon}.

In this context the Ising model with fixed magnetization (IMFM) presents
some very peculiar features. For small systems the microcanonical caloric
curve which relates energy to temperature does not show any backbending in
the temperature domain where a first order phase transition exists in the
Ising model~\cite{richert}. 
This finding shows that backbendings are deeply connected to the constraints 
which are imposed on the system~\cite{richert,spherical} and their presence may
depend on the physical quantities which are 
used in order to define the considered statistical ensemble as
it will be shown later. It is well known that IMFM is isomorphous to
an isochore Lattice Gas model. If we consider that no divergence is
expected for the heat capacity at constant volume in the macroscopic liquid
to gas phase transition, the absence of a backbending in the microcanonical
caloric curve illustrates the fact that there may be
%In particular,there may be
no one to one correspondence between first order phase transitions and
convexity anomalies of the microcanonical entropy as a function of energy. 
%Moreover
However, according to Ref.~\cite{pleimling}, a backbending appears
in the caloric curve of very large lattices. The thermodynamic limit of this
behavior is not clear in view of the present status of Montecarlo
simulations. 
%This seems to contradict the present
%understanding of backbendings as originating from the contribution of
%surface entropy, i.e. finite size effects~\cite{gross} which should become
%increasingly smaller when the size of the system increases. 
Moreover, recent analytic arguments~\cite{kastner} lead to the
conclusion that the transition is continuous in the thermodynamic limit. The
cluster properties of small IMFM lattices also show different signs of
critical behavior~\cite{carmona}. Together with the information coming from
the caloric curve, this could suggest that the apparent order of the
transition in the IMFM model changes with the size of the 
system.

In the present paper, we shall try to clarify these issues by using a
definition of phase transitions based on the occurrence of bimodalities in
the probability distribution of an observable~\cite{topology}. 
This definition is a generalization of the one based on curvature anomalies
and satisfies the Yang Lee unit circle theorem in the thermodynamic 
limit~\cite{lee,topology}. Since, in
Gibbs equilibria, a bimodal distribution of an observable is connected to
the backbending of the associated equation of state in the ensemble where
the observable is constrained by a conservation law, this study will allow
us to elucidate the relationship between backbendings, constraints, and the
order of the transition. 
%In this paper we shall try to clarify these issues 
%by using a complementary approach to the 
%definition and study of phase transitions in finite 
%systems based on the topology of the event 
%distributions in the observable space\CITE{topology}. 
%In this approach a first order phase transition is 
%defined as a bimodality in the probability 
%distribution of an observable which can 
%then be seen as an order parameter. This 
%definition can be linked to the distribution 
%of zeroes of the partition sum and satisfies 
%the Yang-Lee unit circle theorem in the 
%thermodynamic limit\CITE{lee,topology}. 
%Moreover, in Gibbs equilibria, a bimodal 
%distribution of an observable is connected 
%to the backbending of the associated equation 
%of state in the ensemble where the observable 
%is constrained by a conservation law. This will 
%allow us to clarify the relationship between 
%backbendings, constraints, and the order of the transition.
%       

We shall first analyze the topology of the events in the Ising model
(section~\ref{sec2}) to show that many properties of IMFM can be 
understood starting from the standard Ising model. 
We shall see that backbendings are not
characteristic of the caloric curve only but can appear in other equations
of state and closely depend on the constraints applied to the system, i.e.
on the statistical ensemble in which the system is studied. In 
section~\ref{sec3} we shall consider the behavior of the IMFM and we shall see 
that for all sizes the transition can be unambiguously recognized as a first 
order transition even if a backbending appears in the microcanonical caloric 
curve only for large sizes and may disappear at the thermodynamic limit.  
Concerning this limit, a first order transition does not imply the convergence 
of the backbending to a finite energy jump.  %latent heat
Some general arguments for a zero  energy discontinuity 
will be given in section~\ref{sec4} and complemented by a
study of cluster properties. In order to confirm the outcome of the
numerical simulations, an analytical model of a finite system with negative
heat capacity converging to a caloric curve which does not show a plateau 
in the thermodynamic limit even if the system is undergoing a 
first order phase transition will be given in section~\ref{sec5}.

\section{Topology of events in the Ising model}
\label{sec2}

We shall first concentrate on the well known Ising model to show that the
topological properties of observable distributions~\cite{topology} allow to
reconstruct the thermodynamics of the model and at the same time to get some
hint about the behavior of the IMFM model. The order parameter of the
standard Ising model is the magnetization $M=\sum_{i}s_{i}$, where $s_{i}=%
\pm 1$ and the sum extends to the $N$ sites of a lattice. It is interesting
to note that because of the mapping between the Ising model and the Lattice
Gas model $s_{i}=2n_{i}-1$, a spin up $s_{i}=1$ can also been interpreted as
an occupied site $n_{i}=1.$  Then the magnetization is mapped into the order
parameter for the Lattice Gas model which is nothing but the number of
particles $A=\sum_{i}n_{i}$.  Since the volume is fixed, $A$ indeed
corresponds to the density $\rho =A/N$, the known order parameter of
the liquid-gas phase transition. We shall also use the average
magnetization per site $m=M/N$ which is then isomorphous to a particle 
density via $m=2\rho -1$. 
In the Ising model neighboring sites interact via a constant attractive
coupling $\epsilon$. In our numerical implementations of the model we have
considered three dimensional cubic lattices characterized by a linear
dimension $L=N^{1/3}$ with periodic boundary conditions. Details about the
Metropolis simulations can be found in Ref.~\cite{carmona_npa}.

\begin{figure}[htbp]
\begin{center}
\resizebox{18cm}{!}{\includegraphics
%[height=0.5\textheight,trim = 100 250 100 50]%
{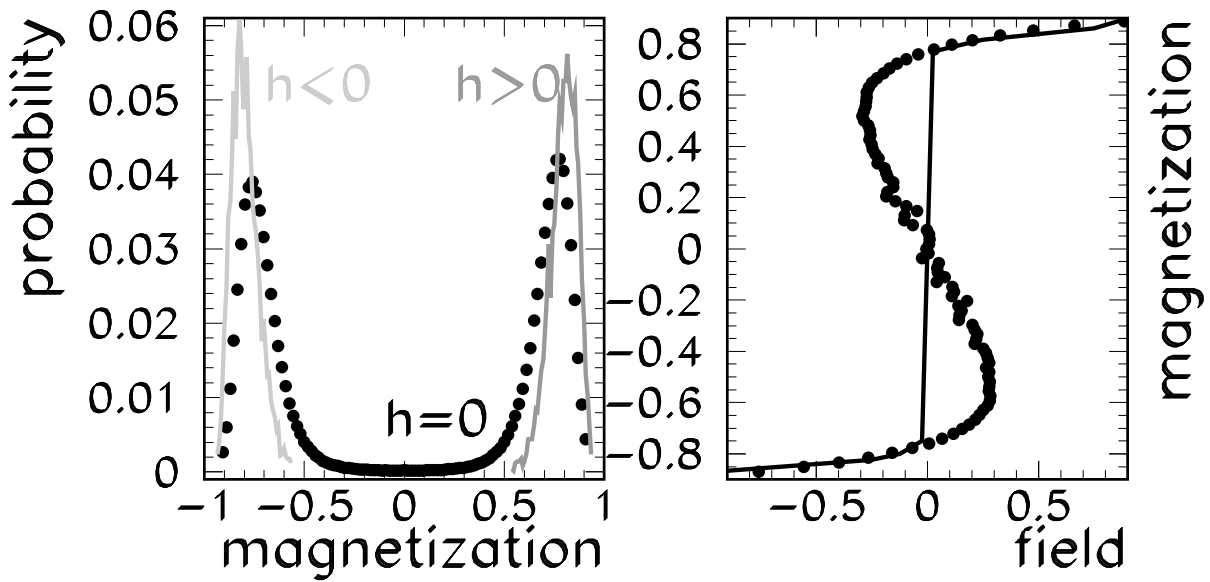}}
\end{center}
\caption{Ising calculation of an $L=6$ cubic lattice with periodic boundary
conditions at a subcritical temperature $T < T_c$. Left side: magnetization
distribution with a negative, zero and positive external field. Right side:
(full line) average magnetization as a function of the external field;
(symbols) field $h_M$ in the IMFM model obtained by derivation of the 
magnetization distribution at zero field [see Eq.~(\ref{eos})].}
\label{fig1}
\end{figure}

\begin{figure}[ht]
\begin{center}
\epsfig{figure=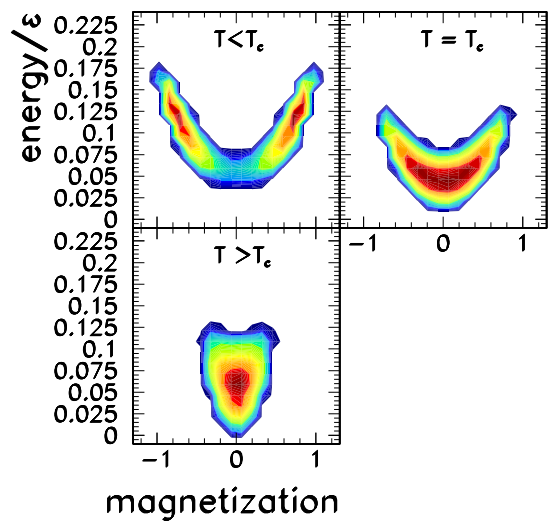,height=10.cm} \vspace{-1.cm}
\end{center}
\caption{Magnetization versus energy distributions with a zero external
field for an $L=6$ Ising lattice at different temperatures.}
\label{fig2}
\end{figure}

Below the critical temperature $T_{c}$ the Ising model shows a first order
phase transition at zero field $h=0$ which is characterized, 
in infinite systems, by a discontinuity of the
magnetization when the field changes sign. This jump is clearly visible in
the standard average magnetization versus field equation of state already
for a lattice as small as $L=6$ as shown in the right part of
Fig.~\ref{fig1}. It is important to notice that for a finite system
the magnetization is an observable which does not take a unique value but 
fluctuates from event to event so that the constant field statistical
ensemble corresponds to a whole magnetization distribution. 
This distribution is shown in the left part of Fig.~\ref{fig1} 
for three different values of the
applied external field. For sufficiently large fields $h<0$ ($h>0$) 
the system gets a negative (positive) magnetization peak. The
transition from the dominantly negative to the dominantly positive solution
passes through a bimodality of the magnetization distribution which can be
taken as a definition of phase coexistence, the two peaks being associated
to the two phases~\cite{topology,lee,binder}. The transition field is the
one for which the two peaks have the same height so that the most
probable magnetization jumps from negative to positive values.
For symmetry reasons this of course occurs at $h=0$.   
The intuitive understanding of phase
coexistence given by this topological definition is clearly seen in 
Fig.~\ref{fig2} which gives the magnetization distribution as a 
function of temperature for the same $L=6$ lattice at $h=0$ field. 
At supercritical temperatures the
distributions are normal while the critical point can be seen as a
bifurcation point from which the distribution splits into two separated
components or phases.

It is important to notice that a bimodality in the distribution of an
observable (here the magnetization) implies a backbending for the associated
equation of state in the statistical ensemble where this observable is
controlled on an event by event basis. Indeed the magnetization distribution
of the Ising model in the presence of a fixed magnetic field at a
temperature $T=\beta^{-1}$ reads 
\begin{equation}
p_{\beta h}(M)=\frac{1}{Z_{\beta h}}\sum_{E}W(E,M)e^{-\beta (E-hM)}=\frac{%
Z_{\beta }(M)}{Z_{\beta h}}e^{(\beta hM)}
\label{p_Ising}
\end{equation}
where $W(E,M)$ is the number of configurations or microstates with energy $E$
and magnetization $M$, and the relation between the partition sum of Ising with
a magnetic field $Z_{\beta h}$ and at fixed magnetization $Z_{\beta}(M)$ is 
\begin{equation}
Z_{\beta h}=\sum_{M}Z_{\beta}(M)e^{(\beta hM)}=\sum_{E,M} W(E,M)
e^{-\beta(E-hM)}.
\label{partition_sums}
\end{equation}
Equation~(\ref{p_Ising}) shows that the magnetization distribution 
$p_{\beta h}(M)$ in standard Ising can be directly related to the partition 
sum of the IMFM. The thermodynamic relations for the IMFM can thus at least in
principle be calculated from $p_{\beta h}(M)$ without a direct simulation of
the IMFM. In particular the equation of state related to the magnetization
in the IMFM reads 
\begin{equation}
h_{M}\equiv-\frac{1}{\beta }\frac{\partial \log Z_{\beta}(M)}{\partial M}=
-\frac{1}{\beta }\frac{\partial \log p_{\beta h}(M)}{\partial M}+h. 
\label{eos}
\end{equation}
Equation~(\ref{eos}) shows that a bimodality in $p_{\beta h}(M)$ implies a
backbending of the intensive parameter $h_{M}$ (which has the dimension of a
magnetic field) associated with the magnetization as shown on the r.h.s.
of Fig.~\ref{fig1}. 
It is important to stress that $h_{M}$ is not a mathematical
artifact but represents the physical magnetic field which, applied to Ising,
gives $M$ as the most probable response. Indeed in the presence of an
external field $h$, the most probable magnetization should fulfil 
\begin{equation}
\frac{\partial \log p_{\beta h}(M)}{\partial M}=\frac{\partial \log Z_{\beta
}(M)}{\partial M}+\beta h=\beta (h-h_{M})=0  \label{extrema}
\end{equation}
i.e. the (most probable) response is $M$ if the applied field is $h=h_{M}$. 
This is true as long as the system is not undergoing a phase transition 
i.e. as long as Eq.~(\ref{extrema}) has only one solution. Indeed, 
Eq.~(\ref{extrema}) defines the extrema of the probability distribution. 
When the solution is unique this unique extremum can only be a
maximum. In the transition regions i.e. in the backbending region 
Eq.~(\ref{extrema}) has three solutions, two maxima and a minimum in between. 
%In this region eq.(\ref{extrema}) has three solutions, 
%meaning that each value of the applied field is 
%associated with two possible solutions for the 
%magnetization or peaks of the $M$ distribution 
%(and a minimum in between). 
This corresponds to the  subcritical temperatures for which $p_{\beta h}(M)$ is
bimodal in a region around $h=0$. In this regime the interval $\Delta h_{M}$
associated with the backbending corresponds exactly to the interval $\Delta h
$ for which $p_{\beta h}(M)$ is bimodal
\footnote{The values of the magnetization between the two peaks of the probability
distribution are strongly suppressed (see Fig.1), and this
suppression increases with the size of the system. This means that in
practical simulations of large lattices the correspondence between Ising and
IMFM breaks down and Eq.(\protect\ref{p_Ising}) is no longer an efficient method to
calculate the thermodynamic properties of IMFM.}

\begin{figure}[htbp]
\begin{center}
\includegraphics[height=0.5\textheight]{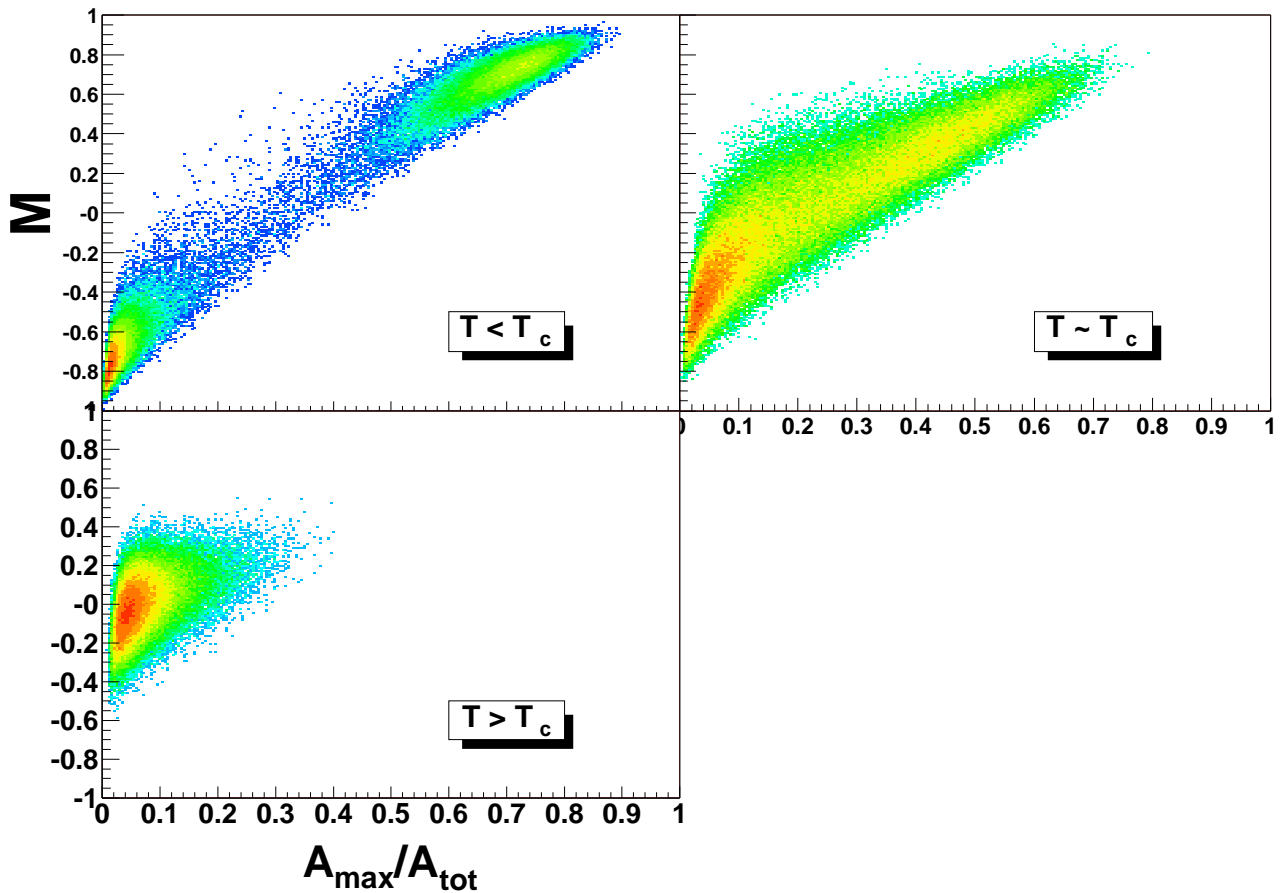}
\end{center}
\par
%$\beta^{-1}=4\epsilon, 4.8\epsilon, 7.2\epsilon$.
\caption{Correlation between magnetization and the size of the largest
connected domain for an $L=6$ Ising lattice with zero field at three
different temperatures $T=4\epsilon, 4.8\epsilon, 7.2\epsilon$.}
\label{fig3}
\end{figure}

This discussion can be immediately extended to any generic Gibbs 
equilibrium~\cite{topology}. 
A phase coexistence (first order phase transition) is
signalled by a bimodality in the probability distribution of an extensive
variable which can then be identified as an order parameter. This
topological anomaly is equivalent to a convexity anomaly in the
thermodynamic potential of the ensemble in which the order parameter is
constrained by a conservation law (extensive ensemble), and reflects in a
backbending of the associated intensive variable in this extensive ensemble.
A backbending in the order parameter equation of state is an alternative
possible definition of a first order phase transition in an extensive
ensemble of a finite system~\cite{gross}.

This observation has a very important consequence. The two definitions of
first order phase transitions in finite systems (bimodalities and
backbendings) are strictly equivalent~\cite{topology}. The observation of a
bimodal probability distribution in an ensemble where the order parameter is
not strictly fixed %controlled 
allows then to conclude about the order of the phase transition in the
corresponding extensive ensemble without explicitly observing or simulating
it.

In particular the bimodality of the magnetization distribution observed for
the standard Ising model with no constraints on the magnetization ($h=0$) at
subcritical temperatures (Fig.~\ref{fig2}) implies that the magnetic
susceptibility of IMFM is negative for magnetization domains lying between the
two dominant peaks of Ising. This negative susceptibility, analogously to a
negative heat capacity in a microcanonical ensemble, indicates a first
order phase transition. The connection between the bimodality of the
magnetization distribution and a negative susceptibility is valid for all 
finite sizes up to the thermodynamic limit. Therefore, the well known fact 
that the Ising
bimodality converges to a finite jump in the thermodynamic limit guarantees
that the corresponding phase transition in IMFM is also first order up to
the thermodynamic limit.
This situation is analogous to the relationship between the grancanonical and
the canonical ensemble discussed in the context of the Yang-Lee
theorems~\cite{vanhove}. Indeed because of the exact mapping between Ising and
Lattice Gas the magnetization can be viewed as a particle density and $\beta$
times the magnetic field is isomorphous to the logarithm of a fugacity.
Using the fact that the equation of state at the thermodynamic limit is the
limiting curve of an analytical function, the equivalence between ensembles can
be demonstrated for short range interactions 
even in the coexistence region of first order phase transitions~\cite{vanhove}.

\begin{figure}[tbph]
\begin{center}
\resizebox{18cm}{!}{\includegraphics
%[height=0.8\textheight]%
{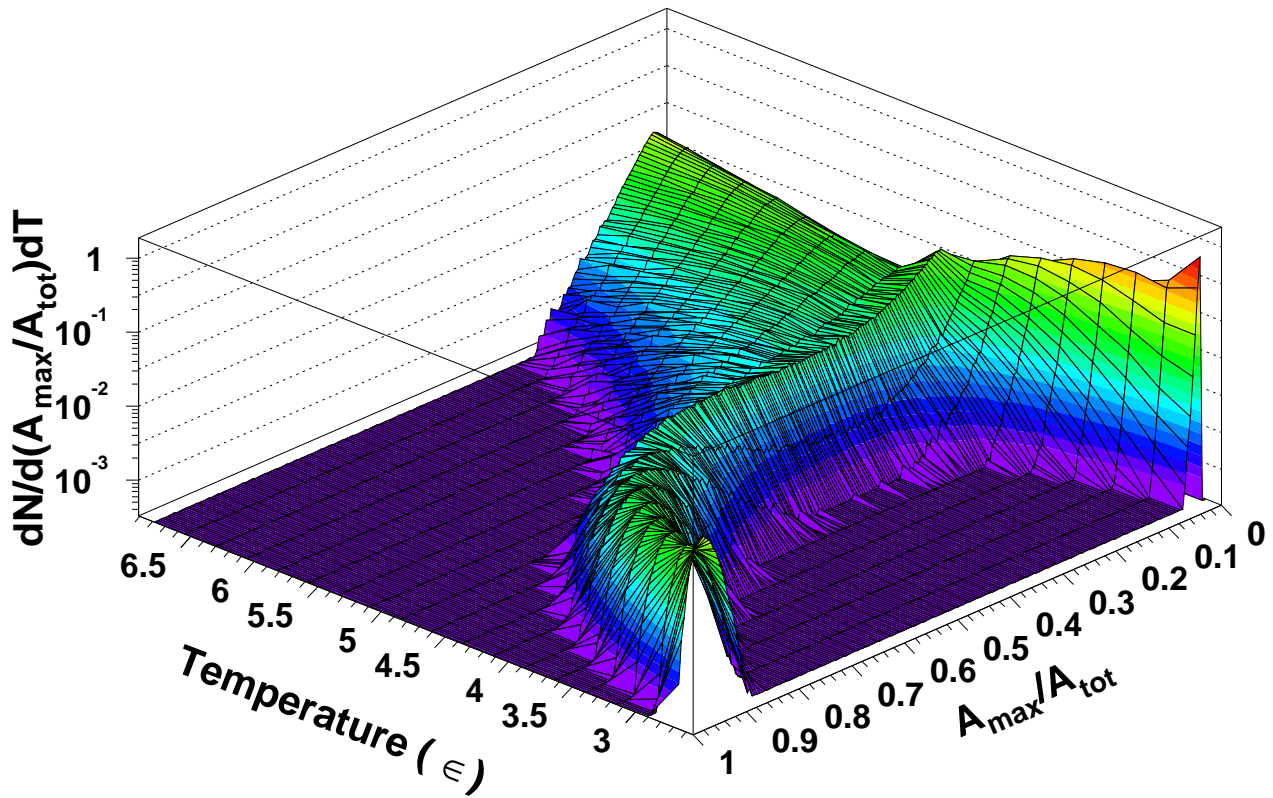}}
\end{center}
\caption{Distribution of events at zero field as a function of the size of
the largest connected domain $A_{max}$ and the temperature $T$, for an $L=6$
Ising lattice.}
\label{fig4}
\end{figure}

In the recent literature of phase transitions in finite systems~\cite{gross}
backbendings have been often observed in the microcanonical caloric curve,
leading to negative heat capacity. Following the general relationship
between ensembles discussed above, a negative heat capacity univocally
implies a bimodal energy distribution in the corresponding canonical
ensemble. %Indeed we shall show in the next section  
%that canonical simulations of the IMFM model are able to 
%show the microcanonical behaviour of the system.
Figure~\ref{fig2} shows that because of the symmetry between 
spins up and down in the Ising Hamiltonian, 
the two magnetization solutions correspond to the same energy. This means that 
the energy cannot be used as an order parameter for this model. 
%and that the microcanonical caloric curve cannot present a backbending
On the other hand, all variables correlated to the magnetization can
present a bimodality, i.e. can be considered as order parameters. As an
example, Figs.~\ref{fig3} and~\ref{fig4} show the correlation
between the magnetization and
the size of the largest connected domain $A_{max}$, here defined with the
Coniglio-Klein algorithm~\cite{CK} in its simplified version proposed by X.
Campi and H. Krivine~\cite{campi}. These figures also illustrate the
fact that a first order phase
transition in a finite system can be seen as a bifurcation in the
observable plane defined by the control variable (here the temperature) and
(one of) the order parameter(s).

\section{The IMFM and the effect of constraints}
\label{sec3}

In the previous section we have recalled that an order parameter of a first
order phase transition can be defined as any observable $O$ which allows to
separate the two phases, i.e. such that if events are sorted as a function
of $O$, they split into two components separated by a minimum of the 
distribution function. This definition is valid only if the order parameter is
free to fluctuate. If the order parameter is constrained by a
conservation law, the first order character of the transition can still be
recognized from the backbending of the equation of state relating the order
parameter to its conjugated intensive variable. For example, a first order
phase transition in a microcanonical model is signalled by a backbending in
the caloric curve. Some models have been reported where the kinetic energy
plays the role of an additional order parameter and shows a bimodal
distribution in the microcanonical ensemble~\cite{lynden,inequivalence} but
in the general case bimodalities have not to be expected if the order
parameter is fixed by a conservation law. In the
same way since the magnetization is the same for all events in the IMFM
model, the Ising bimodality at zero field observed in the previous section
cannot subsist any more in IMFM. 
%Moreover at the thermodynamic limit $M$ and 
%$A_{max}$ follow the same scaling law, meaning that in this limit $A_{max}$
%cannot be bimodal either. 
However in the case of finite systems fluctuations are not negligible and
the relationship between $M$ and $A_{max}$ is not a one to one
correspondence but rather a %large
loose correlation. This implies that for finite systems a bimodality in the 
$A_{max}$ distribution can remain even in the constant magnetization ensemble.
\begin{figure}[tbph]
\begin{center}
\resizebox{18cm}{!}{\includegraphics
%[height=0.8\textheight]%
{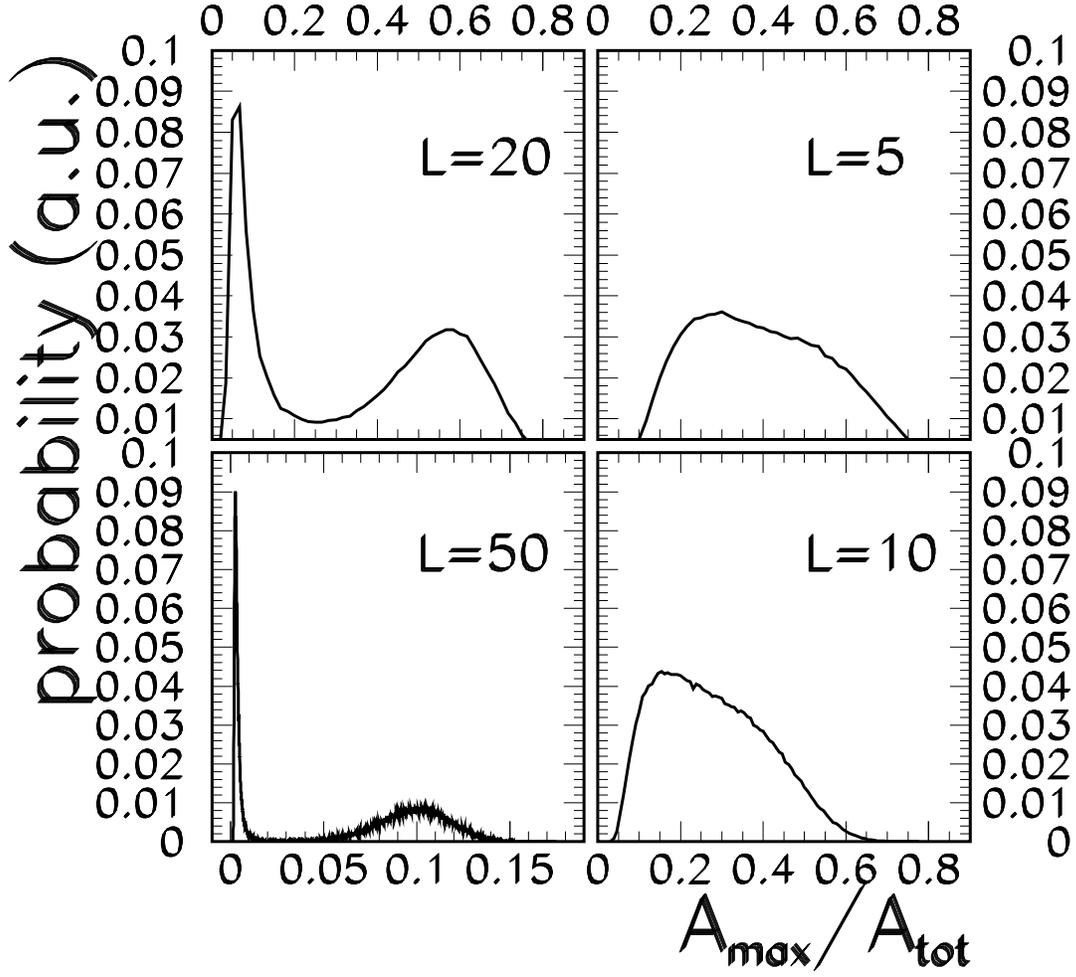}}
\end{center}
\caption{IMFM largest connected domain distribution for: (upper part)
lattice sizes $L=20$ and $L=5$ with the same $A=60$ number of positive
spins; (lower part) lattice sizes $L=50$ and $L=10$ with the same density of
positive spins $\rho =0.13$ (same magnetization $m_{0}=-0.74$).}
\label{fig5}
\end{figure}
This is shown in Fig.~\ref{fig5}, which displays the $A_{max}$ distribution for
the IMFM with a number $A=60$ of positive spins (or Lattice Gas particles) in a
$L=20$ cubic lattice (corresponding to a magnetization $m=M/N=-0.985$  
and a low density $\rho=0.0075$) at a temperature $T=T_{tr}$ where 
$T_{tr}$ corresponds to the transition temperature, i.e. the temperature
at which the two maxima of the energy distribution have the same height. As
shown in Fig.~\ref{fig6} (left), 
$A_{max}$ is correlated with the energy $E$, so
that the bimodality in $A_{max}$ reflects in a bimodality in the energy
distribution. The bimodality in the energy distribution implies that the
microcanonical IMFM caloric curve presents a backbending. We have just
interpreted this backbending as a finite size effect (and we shall come back
to this point in the next section). However for very small systems this
behavior is not visible. This is shown in the right parts of Figs.~\ref{fig5} 
and~\ref{fig6} below, in agreement with the results of Ref.~\cite{pleimling}. 
In the upper right part of Fig.~\ref{fig5}, 
the same system of $A=60$ positive spins
now occupies an $L=5$ lattice, corresponding to a magnetization $m=-0.04$ 
(or a density  $\rho=0.48$). The calculation has been
done at a temperature such that the average energy $\langle E\rangle$ 
is the same as in the $L=20$ case, to be sure that the expected region of the 
backbending is explored. The $A_{max}$ as well as the $E$ distributions are 
monomodal, and no bimodality is ever seen for any value of $T$ in this small 
lattice. The fact that this is correlated to a finite size effect is 
demonstrated in the lower part of Fig.~\ref{fig5}, 
where the same trend is observed keeping the reduced
magnetization constant and varying only the size of the lattice.

\begin{figure}[ht]
\begin{center}
\resizebox{12cm}{!}{\includegraphics
%[height=0.8\textheight]%
{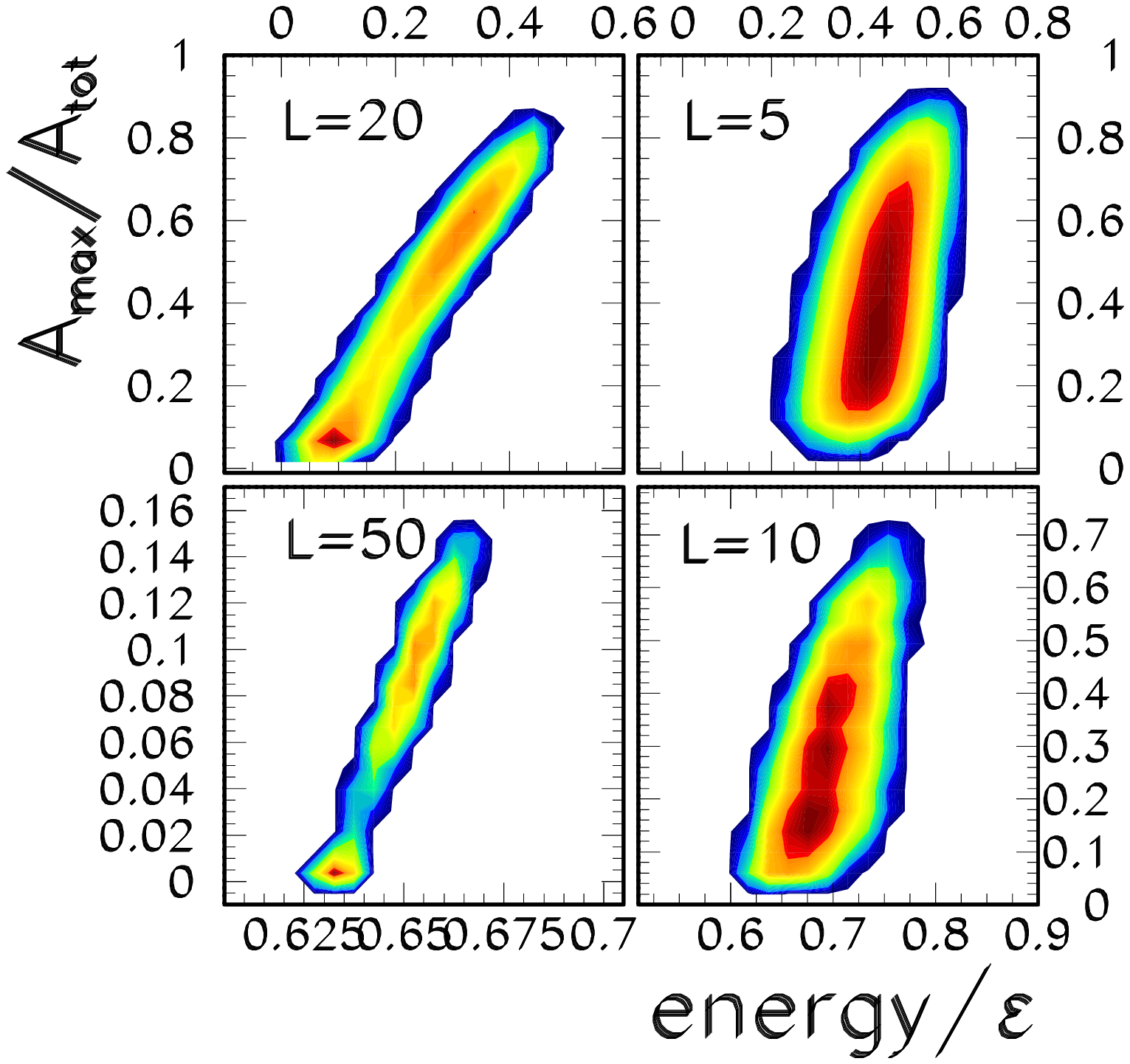}}
\end{center}
\caption{$A_{max}$ versus energy correlations for: (upper part) an $A=60$
system in an $L=20$ and $L=5$ lattice; (lower part) a $\rho=0.13$ system in
an $L=50$ and $L=10$ lattice.}
\label{fig6}
\end{figure}

This behavior can be understood because the correlation between $A_{max}$
(and so the energy) and $M$ becomes %wider 
looser the smaller the size of the system. For very small systems the width
of the two peaks becomes comparable to their distance and the bimodality
cannot be seen any more.

\begin{figure}[ht]
\begin{center}
\epsfig{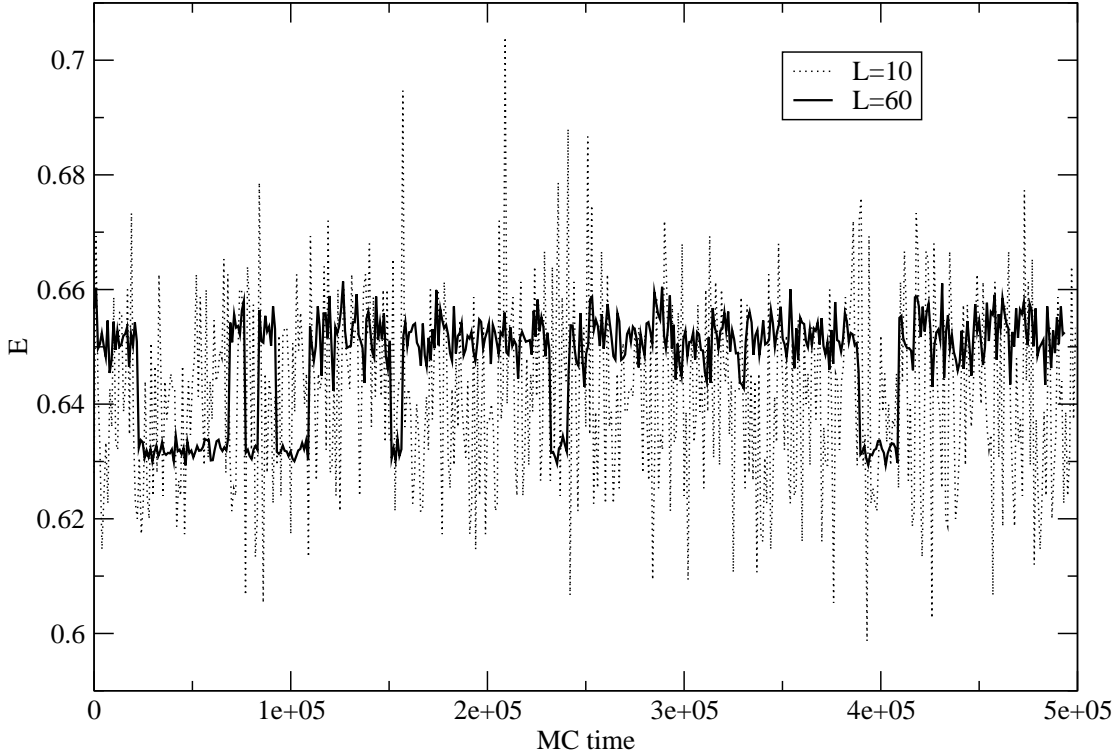}
\end{center}
\caption{Total energy as a function of the Metropolis step for $m=0.75$ and
lattices of linear size $L=10$ and $L=60$ at the transition temperature.}
\label{fig7}
\end{figure}

An intuitive understanding of this phenomenon can be obtained looking at
Fig.~\ref{fig7}, which gives the total energy as a function of the
Metropolis steps for 
%the two systems in the lower part of figure~\ref{fig6}.
two systems, $L=10$ and $L=60$, at the same density $\rho=0.13$.
The width of the distribution is comparable for the two lattice sizes (the 
energy jump  %latent heat 
evolves slowly with $L$~\cite{pleimling}), but in the small lattice the
large fluctuations around each solution do not allow to resolve the two
peaks~\cite{calvo}. An equivalent way of expressing the same idea is to
recall that the two different energy solutions or phases which give the
backbending correspond to very different spatial extensions. These
configurations cannot be explored in the small lattice because of the
boundary condition constraint.

\section{Size distributions and the thermodynamic limit}
\label{sec4}

We have just shown that for (not too) small systems, the energy distribution
of the IMFM can be bimodal. A systematic study of this phenomenon can been
found in Ref.~\cite{pleimling}. If this bimodality (or, equivalently,
backbending of the microcanonical caloric curve) would survive up to the
thermodynamic limit, at the canonical temperature $T_{tr}$ at which the two
peaks have the same height, one would observe at the thermodynamic limit a
jump in the average energy from the disordered to the ordered phase. $T_{tr}$
would then be the transition temperature of a conventional first order phase
transition with a finite latent heat. 
\begin{figure}[htbp]
\begin{center}
\resizebox{18cm}{!}{\includegraphics
%[height=0.8\textheight]%
{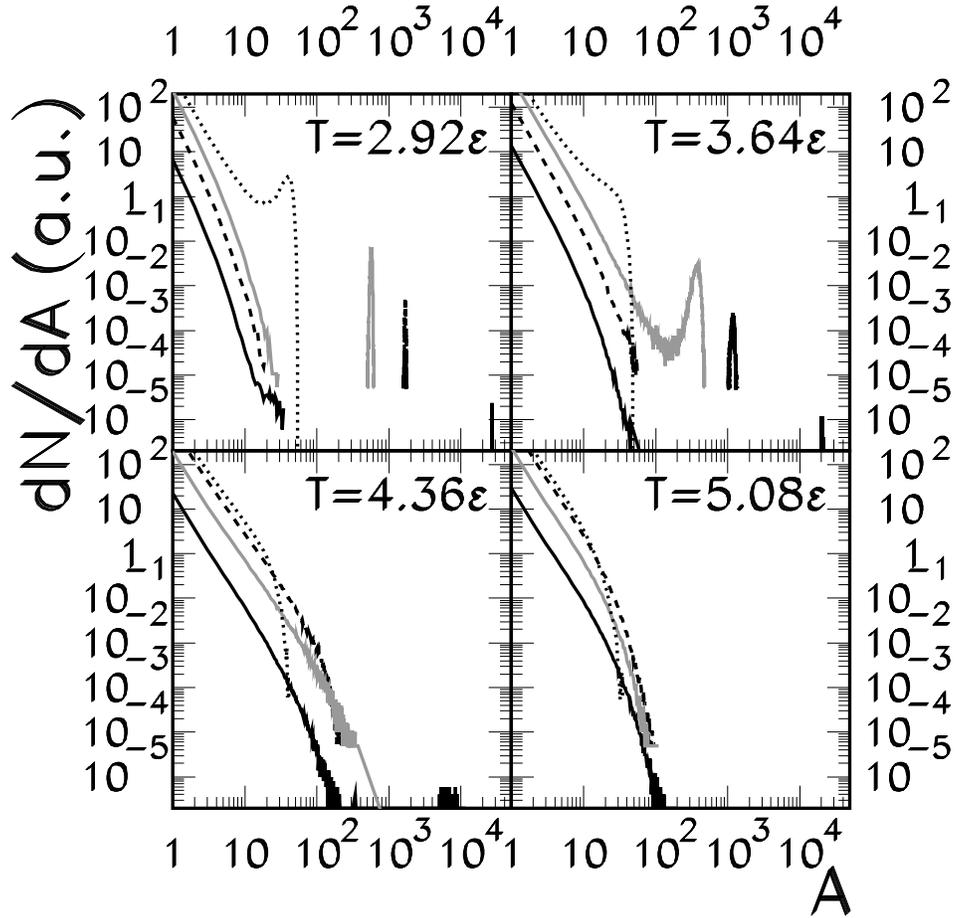}}
\end{center}
\caption{Distribution of cluster size in the IMFM at four different
temperatures and a fixed magnetization density $m=-0.5$ for a cubic lattice
of linear dimension $L=50$ (black) $L=20$ (dashed), $L=14$ (grey)
and $L=8$ (dotted).}
\label{fig8}
\end{figure}
In section~\ref{sec3} we have noticed that the bimodality in energy is 
correlated to a bimodality in $A_{max}$, therefore an indication about
the thermodynamic limit of the energy distribution can be obtained
from an inspection of the size distribution.

\begin{figure}[tbph]
\begin{center}
\resizebox{18cm}{!}{\includegraphics
%[height=0.5\textheight]%
{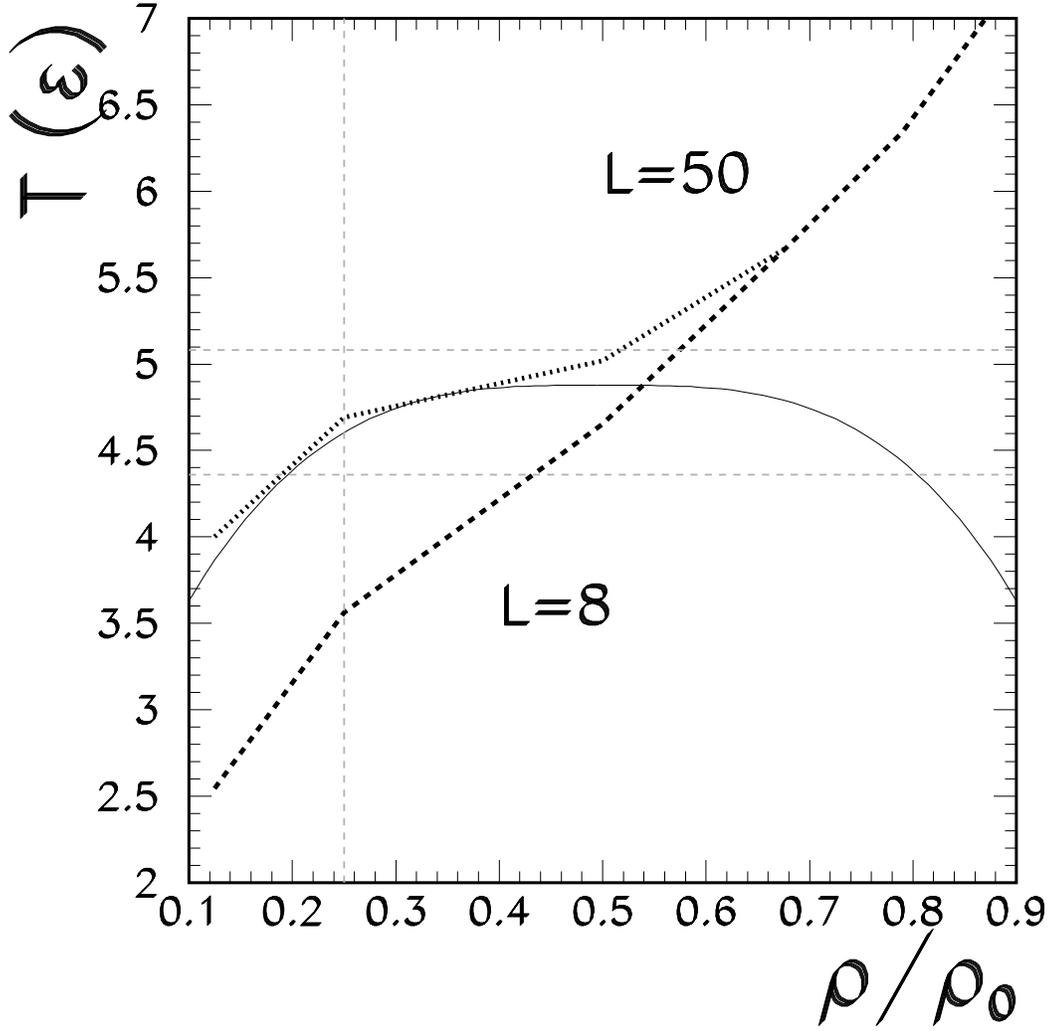}}
\end{center}
\caption{Phase diagram of the IMFM with periodic boundary conditions. Full
line: coexistence line from Refs.~\cite{prl99}; dashed (dotted) line:
curve of the minimum chi-square two parameter fit of the size distribution
with a power law for $L=8$ ($L=50$).}
\label{fig9}
\end{figure}

Figure~\ref{fig8} shows the distributions of fragment sizes at the fixed
magnetization density $m=-0.5$. The corresponding phase diagram is reported
in Fig.~\ref{fig9}~\cite{prl99}. At a given lattice size, a two parameter power
law fit has been performed for each temperature; the temperature at which the
$\chi^2$ of the fit is minimum is then defined as the ``critical'' 
temperature and
reported in Fig.~\ref{fig9} as a dashed line (for $L=8$) and as a dotted line
(for $L=50$)~\cite{prl99}.
A first order transition is clearly
indicated in  the bigger lattice calculation. 
At a temperature slightly lower
than $T_{c}$ (lower left panel of Fig.~\ref{fig8} and lower horizontal
line in Fig.~\ref{fig9}) the ``infinite'' percolation cluster is still present.
At a temperature slightly higher than $T_{c}$ (lower right panel of 
Fig.~\ref{fig8} and higher horizontal line in Fig.~\ref{fig9}) 
the ``infinite'' percolation cluster has disappeared. As it can be seen 
from Fig.~\ref{fig8}, the disappearance point of the
percolation cluster does not correspond to a power law (and in fact finite
size scaling of the cluster size distributions is violated on the dotted
line up to the critical point~\cite{prl99}). Similar results have also been
presented in Ref.~\cite{carmona}. 
In the case of very small lattices the percolation cluster is so small that
its size is comparable to the size of the other fragments; this creates an
apparent and transient power law behavior in the middle of the coexistence
zone~\cite{prl99} which is a finite size effect which should not be confused
with a continuous transition. The sudden disappearance of the percolation
cluster suggests a finite jump in the distribution of $A_{max}$ at the
transition temperature in the thermodynamic limit, and consequently a finite
latent heat. For the results of Fig.~\ref{fig8} to be compatible with zero 
energy jump %latent heat 
it is however enough that the size of the largest cluster increases more 
slowly than $L^{3}$ with increasing lattice size.

From a conceptual point of view we do not expect that the bimodality of the 
$A_{max}$ distribution or equivalently the backbending of the caloric curve
converge to a discontinuity  at the thermodynamic limit. 
Indeed the Van Hove theorem~\cite{vanhove} guarantees, for
short range forces, the equivalence of the ensembles at the thermodynamic
limit. 
Therefore the IMFM equations of state should converge to the Ising
ones even in the coexistence region of first order phase transitions.
 Moreover the Ising model belongs to the liquid-gas universality class  
(the IMFM is in fact isomorphous to the canonical lattice-gas model at
constant volume) and this implies that the IMFM at the thermodynamic limit
should be equivalent to the macroscopic liquid gas phase transition at
constant volume which exhibits a continuous caloric curve with no 
energy jump at a constant temperature.  %(i.e. with no latent heat)
On the other side, if one constructs a Lattice Gas model at fixed 
pressure~\cite{iso-iso} the backbending is visible even for very small
lattice sizes and should converge to a plateau 
in the thermodynamic limit. The constant
volume and constant pressure situations are the same in the particular case
of zero pressure which corresponds to an infinite volume independent of the
number of particles (i.e. of the magnetization $M$). Since the infinite
volume is a constant pressure situation, a backbending is expected for the
microcanonical caloric curve of the finite system in an infinite volume 
i.e. for a density going to zero. This is indeed what is observed in 
Fig.~\ref{fig6} above. The left side of Fig.~\ref{fig6} can be
interpreted in this context as a calculation at constant %(zero) 
(low) pressure, such that the spatial extension of the system is not
constrained by any boundary conditions ($N\rightarrow \infty $). Within this
interpretation it is natural that the backbendings observed in 
Ref.~\cite{pleimling} appear for large lattice sizes compared to the number of
particles. Going towards
the thermodynamic limit however the isobar and isochore paths correspond to
different physical situations and an energy jump %non zero latent heat 
should be associated only to a transformation at constant pressure.

\begin{figure}[htbp]
\begin{center}
\includegraphics[height=0.5\textheight]{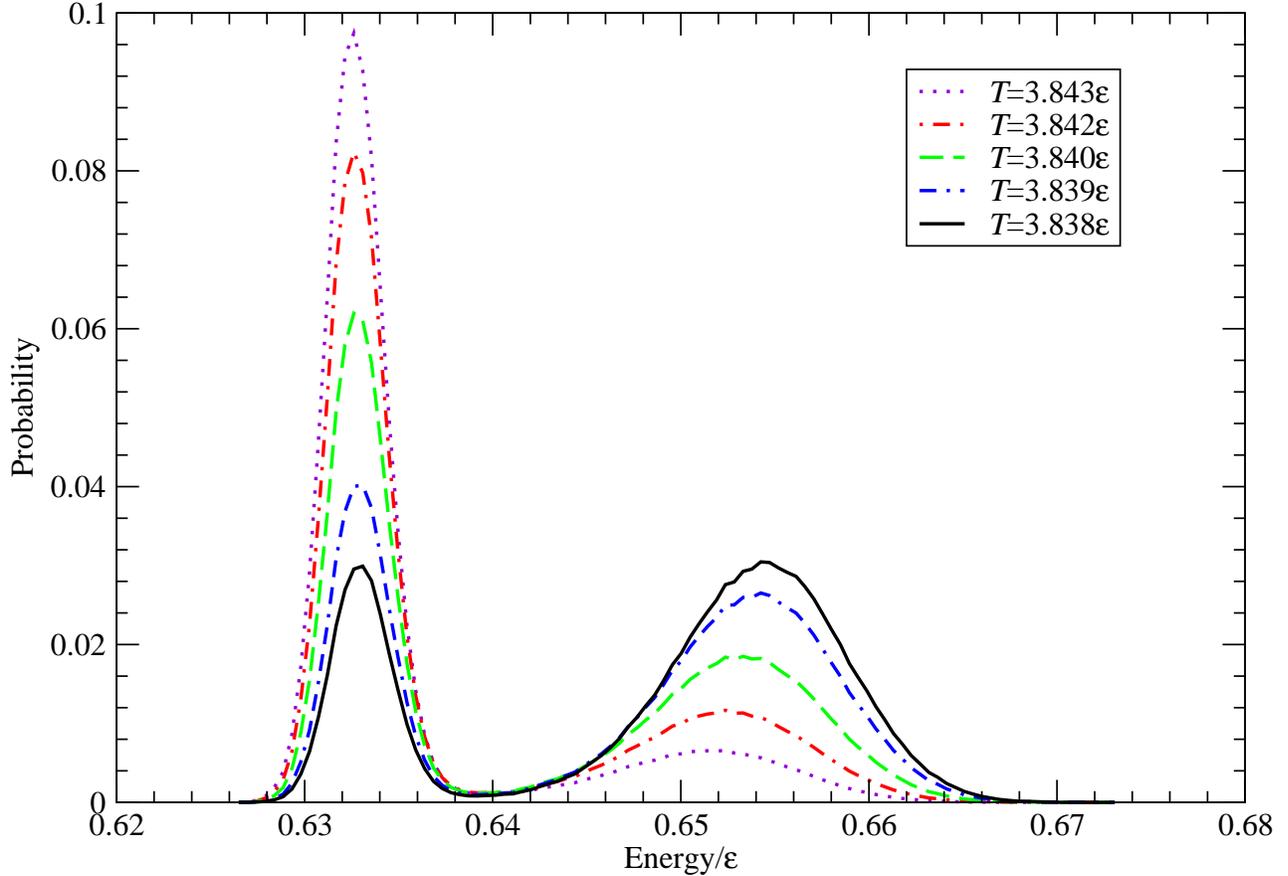}
\end{center}
\caption{Ferrenberg-Swendsen calculation of energy histograms for an $L=50$
lattice at $\rho=0.13$. $T_{tr}$ is found to be $3.838\epsilon$.}
\label{fig10}
\end{figure}

The transient nature of the bimodality can be in principle directly
demonstrated looking at the distance between the two energy peaks at the
transition temperature (energy jump%latent heat 
)~\cite{pleimling} at fixed magnetization as a function of the lattice size.
%Since the bimodality is not seen for small lattices, $L\lesssim 40$,
%this comparison has to be done for rather large lattice sizes. 
Table~\ref{table} shows the magnitude of the energy jump between the two
peaks for several lattices between $L=42$ and $L=60$ at $\rho=0.13$.
This calculation is excessively delicate.
%involves several difficulties.
%Since the energy distance between the peaks changes with the system
%temperature, one has to give a functional definition of the $T_{tr}$: 
%this will be the temperature at which the two energy peaks
%have the same height. In addition, 
Indeed, the transition region in large lattices
is very narrow, and the bimodality is only seen in a very short range of
$T$ ($\Delta T\sim 10^{-3}\epsilon$ for $L=60$). The transition
temperature is then obtained by a Ferrenberg-Swendsen algorithm~\cite{FS}
which moves the energy histograms found at a certain $T$ at which 
the bimodality is present, and finds the $T_{tr}$ at which both peaks have
the same height. Figure~\ref{fig10} illustrates this procedure.

\begin{table}[tbp]
\begin{tabular}{@{\extracolsep{0.5cm}}ccccc}
 & $L$ & $T_{tr}(\epsilon)$ & $\Delta E/\epsilon$ & \\ \hline \hline 
 & 42 & 3.807(1) & 0.0219(1) & \\ \hline 
 & 46 & 3.824(1) & 0.0216(2) & \\ \hline 
 & 50 & 3.838(1) & 0.0212(1) & \\ \hline 
 & 54 & 3.851(1) & 0.0210(1) & \\ \hline 
 & 60 & 3.866(1) & 0.0200(2) & \\ \hline 
\end{tabular}
\caption{Transition temperatures and difference between the two 
(intensive) energy values of the peaks present in the energy histogram
of several lattice sizes between $L=42$ and $L=60$ at $\rho=0.13$.}
\label{table}
\end{table}

The results of Table~\ref{table} show a tendency of the distance
between the peaks to decrease, although very slowly. 
Extending this study to larger lattices is a very hard computational task.
Appart from the increase in computer time of Monte Carlo steps ($\propto L$),
when $L$ is very large, the Monte Carlo time between jumps increases too,
what makes very difficult to perform the Ferrenberg-Swendsen analysis 
described above. For the moment, making this analysis for lattice sizes
$L>60$ lies beyond the possibilities of our present computers.

\section{An analytical model}
\label{sec5}

The fact that energy can play the role of an order parameter in the IMFM, as
well as its transient character, has been up to now discussed in the
framework of numerical simulations. Whatever the numerical procedure, the
quality of the exploration of the phase space becomes increasingly delicate 
with increasing number of possible configurations, i.e. size of the lattice. In
this section we shall therefore make use of the fact that Ising belongs to
the liquid-gas universality class to show in a simple analytical liquid drop
model that

\begin{itemize}
\item  the microcanonical caloric curve can backbend at constant volume
\item  this backbending disappears at the thermodynamic limit
\item  the convergence can be very slow and it is not surprising that an
apparent energy jump  %latent heat 
still survives for lattice sizes as large as $L=50$.
\end{itemize}

\begin{figure}[tbph]
\begin{center}
\resizebox{18cm}{!}{\includegraphics
%[height=0.5\textheight,trim = 100 250 100 50]%
{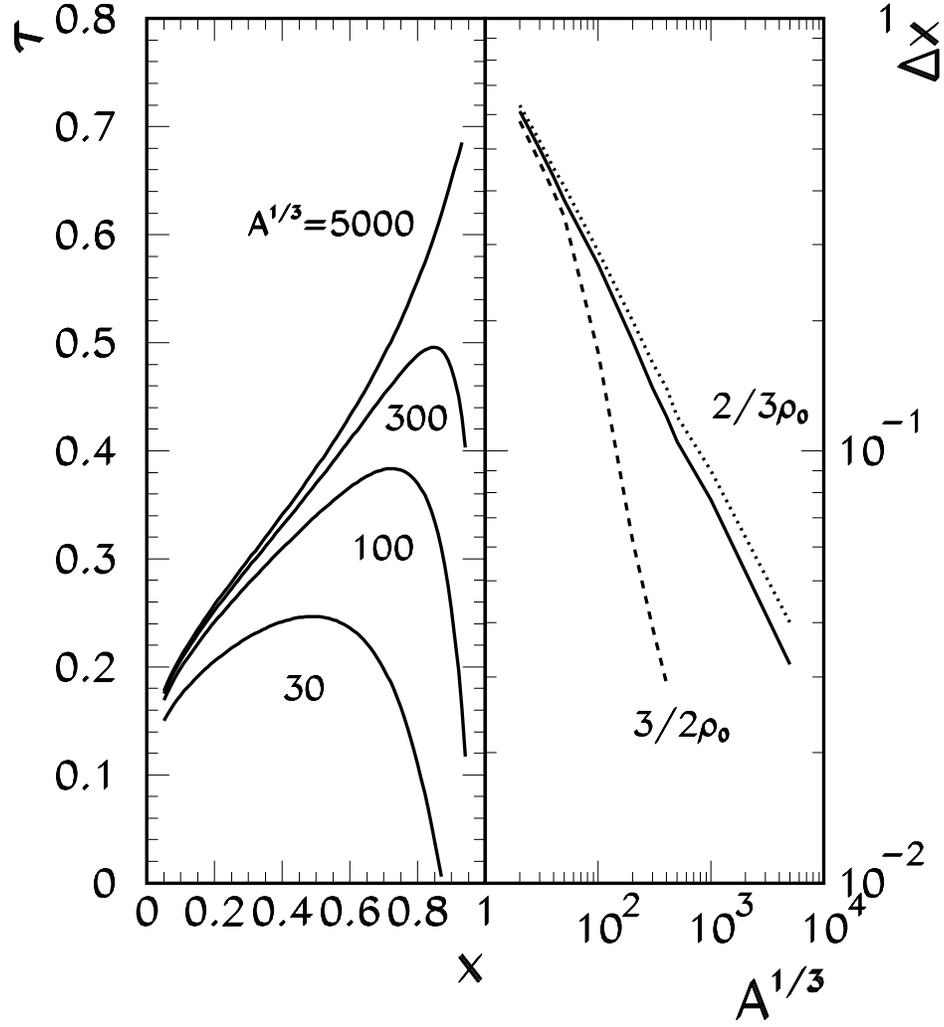}}
\end{center}
\caption{Equation of state of a liquid droplet in equilibrium with its
vapor from Eq.~(\ref{eos_anal}). Left side: reduced temperature as a
function of the vapor fraction for different sizes of the system. Right
side: vapor fraction interval of the negative heat capacity region as a
function of the size of the system. Full line: same total density as in the
calculation of the left panel; dashed (dotted) line: higher (lower)
density.}
\label{fig11}
\end{figure}

Let us consider a liquid-gas phase coexistence as a spherical liquid drop in
equilibrium with an ideal gas of monomers~\cite{moretto}. This picture will
not apply to very small systems (for which the liquid drop approximation as
well as the monomer approximation will fail), while the thermodynamic limit
will be recovered by letting the droplet radius $R\rightarrow \infty$. The
equilibrium between the droplet and the vapor can be obtained by equalizing
the chemical potential of the two phases which leads to the Clapeyron equation 
\begin{equation}
\frac{dp}{dT}=\frac{\Delta s}{\Delta v}=\frac{\Delta e+p\Delta v}{T\Delta v}
\end{equation}
which expresses the relation between the pressure $p$ and temperature $T$ in
the coexistence zone of the first order phase transition. Here $\Delta s$,
$\Delta v$, $\Delta e$ represent the difference in entropy, volume and energy
per particle between the two phases. In the low pressure regime the liquid
specific volume $v_{l}$ is much lower than the gas specific volume, $%
v_{l}\ll v_{g}$, and the vapor can be considered as an ideal gas of monomers 
$pv_{g}=T$ leading to~\cite{ma} 
\begin{equation}
\frac{dp}{dT}=\frac{\left( \Delta e+T\right) p}{T^{2}} .
\end{equation}
If in addition we consider the subcritical regime where the temperature is
much lower than the typical latent heat $T\ll \Delta e$ we can consider the
latent heat as a constant $\Delta e(p,T)\approx \mbox{const.}$ 
and the Clapeyron equation is readily integrated giving 
\begin{equation}
p=p_{0}\exp\left( -\frac{\Delta e}{T}\right).   
\label{clapeyron}
\end{equation}
If the liquid fraction is constituted by a finite drop, its binding energy
per particle $\Delta e$ is reduced with respect to its bulk value 
$\Delta e_{0}$~\cite{moretto}, 
$\Delta e=\Delta e_{0}-3a_{s}v_{l}/R$ where $a_{s}$
is the surface energy coefficient and $R=r_{0}A^{1/3}$ is the drop radius.
The equality between the liquid pressure Eq.~(\ref{clapeyron}) and the vapor
pressure $p=T/v_{g}$ gives 
\begin{equation}
p_{0}\exp\left( -\frac{\Delta e_{0}}{T}+\frac{3a_{s}v_{l}}{r_{0}A^{1/3}T}%
\right) =\frac{T(A_{tot}-A)}{V-Av_{l}}
\end{equation}
where $A$ is the mass number of the droplet and $A_{tot}$ is the total mass
of the system (droplet plus vapor) and a strict mass conservation has been
implemented. If we introduce the vapor fraction $x=1-A/A_{tot}$ and the
reduced temperature $\tau =T/\Delta e_{0}$, the relation between $\tau$ and 
$x$ (which is monotonically correlated to the energy) at constant volume can
be written as 
\begin{equation}
c_{1}-\frac{1}{\tau }+\frac{c_{2}}{\tau (1-x)A_{tot}^{1/3}}=\log \tau +\log 
\frac{x}{1-c_{3}(1-x)}.  
\label{eos_anal}
\end{equation}
Here $c_{1}=\log (p_{0}/\rho \Delta e_{0})$, $c_{2}=3a_{s}v_{l}/r_{0}\Delta %
e_{0}$, $c_{3}=\rho v_{l}$ contain the specific features of the physical
system under study and are linked to its bulk pressure, surface properties
and to the system volume respectively. In the thermodynamic limit $%
A_{tot}\rightarrow \infty $ the relation between the temperature and the
vapor fraction (i.e. the caloric curve) Eq.~(\ref{eos_anal}) is monotonic
in the physical domain $0<x<1$ for all physical values of the constants as
expected. For finite systems the equation of state can present a backbending
with an amplitude depending on the value of the parameters $c_{1},c_{2},c_{3}
$. An example is given in Fig.~\ref{fig11} for $c_{1}=1$, 
$c_{2}=0.25$, $c_{3}=0.1$. 
The quantity $c_{1}$ represents a global shift and does not influence
the monotonic character of the equation of state. The quantity $c_{2}$
governs the speed of convergence towards the thermodynamic limit while the 
influence of $c_{3}$ is shown in the right part of Fig.~\ref{fig11}. 
The backbending progressively decreases with the
increasing size of the system but this phenomenon is not specific of very
small droplets only. In order to quantify the evolution towards the
thermodynamic limit one may study the variation of the extension of the 
backbending region with the size of the system. To this aim we have 
represented the vapor fraction interval corresponding to the
slope inversion of the equation of state in the right part of
Fig.~\ref{fig11}. The monotonous correlation between
the energy and the vapor fraction assures that the backbending in $x$
corresponds to a backbending in energy, i.e. the adimensional quantity 
$\Delta x$ is directly correlated to the energy jump in the backbending region.
%latent heat of the transition
The clear power law behavior as a function of the total mass of the system
shows that the   energy jump  %latent heat 
goes to zero only at the thermodynamic limit. 

\section{Conclusions}
\label{sec6}

In this paper we have analyzed the different effects due to the finite size
of the systems on the determination of order parameters and on the
definition and classification of phase transitions using the Ising model
with fixed magnetization (IMFM). A detailed comparison with the standard
Ising model at zero field has shown the general effect of constraints on the
equations of state of finite systems. Indeed ensembles are in general not
equivalent in finite systems and in particular first order phase transitions
manifest themselves in a very different way in the ensemble in which a
constraint is put on the order parameter or on an observable closely
connected to it~\cite{inequivalence}. If we consider an ensemble in which a
constraint is put on the order parameter,  %like in the case of the IMFM
the bimodality in the order parameter distribution reflects as a
backbending in the equation of state which links the order parameter to its
associated intensive variable. Looking at the global topology of events in
the observables space~\cite{topology} we have shown that the first order
character of the transition also manifests itself even in lattice sizes as
small as $L=6$ through bimodalities in variables only loosely connected to
the magnetization. The connection between the event topology and the cluster
distribution led to the conclusion that the apparent signs of a continuous
transition seen in this model for small lattice
sizes~\cite{prl99,richert,carmona} can be interpreted as a finite
size effect. Indeed fluctuations around the transition temperature can 
lead to an anomolously wide cluster distribution with an apparent power 
law behavior and eventually a bimodality in the distribution of the largest 
connected domain. Because of the correlation between the cluster sizes and 
the total energy we have suggested that the anomalous backbendings observed 
in the caloric curve for very large lattices~\cite{pleimling} are due to the 
same finite size effects, and should disappear in the thermodynamic limit  
as expected from the van Hove theorem. These speculations are comforted
by the analytical study of a macroscopic liquid droplet in equilibrium with
its vapor in a low density and low temperature approximation. The equation
of state at constant volume presents a backbending in the coexistence region
with an energy jump which decreases as a power law towards the 
thermodynamic limit.

\acknowledgments
We thank A. Taranc\'on for many and very useful discussions. 
This work was partially supported by Spanish MCyT FPA2001-1813
research contract. S. Jim\'enez is a DGA fellow.

\end{document}